\documentclass[11pt]{article}
\usepackage{mathptmx}  %% gives ps fonts
\usepackage{amsfonts,amsmath} 

\newfont{\gl}{eufm10 scaled \magstep1} %% gothic fonts

\begin{document}

\noindent{\huge Poisson reduction and branes}
\vskip 0.2 cm
\noindent{\huge in Poisson-Sigma models}
\vskip 0.5 cm
\noindent IV\'AN CALVO and FERNANDO FALCETO \\
\\ Depto. F\'{\i}sica Te\'orica, Univ. Zaragoza, E-50009 Zaragoza, Spain\\
E-mail: {\tt icalvo@unizar.es, falceto@unizar.es}

\vskip 0.3 cm

\noindent{\small Received:}

\vskip 0.3 cm

\noindent {\bf Abstract.} 
We analyse the problem of boundary conditions for the Poisson-Sigma
model and extend previous results showing that non-coisotropic branes
are allowed. We discuss the canonical reduction of a Poisson structure
to a submanifold, leading to a Poisson algebra that generalizes
Dirac's construction. The phase space of the model on the strip is
related to the (generalized) Dirac bracket on the branes through a
dual pair structure.

\vskip 0.3 cm


\vskip 0.3 cm


\vskip 0.5 cm

\section{Introduction}

Poisson-Sigma models (\cite{Strobl}, \cite{Ikeda}) are topological
field theories whose field content is a bundle map from the tangent
bundle of a surface $\Sigma$ to the cotangent bundle of a Poisson
manifold $M$. Their initial interest was due to the fact that some
two-dimensional gauge theories such as pure gravity, WZW models and
Yang Mills are particular cases of Poisson-Sigma models
(maybe after the addition of a non-topological term expressed in terms
of Casimir functions of the Poisson structure on $M$).

The models gained renewed attention with the appearance of
\cite{CaFe}, where it was shown that when $\Sigma$ is a disk the
perturbative path integral expansion (with appropriate boundary
conditions) reproduces the $*$-product introduced by Kontsevich in
\cite{Kon} that gives the deformation quantization of a Poisson
manifold.

More recently, A.S. Cattaneo and G. Felder studied in \cite{CaFe3} the
possible boundary conditions for the Poisson-Sigma model and concluded
that the branes of this model are labeled by the coisotropic
submanifolds of $M$. In this paper we show that more general boundary
conditions are allowed and, in fact, we extend their procedure to the
case in which the base map of the bundle map restricts on $\partial
\Sigma$ to an (almost) arbitrary submanifold of $M$.

We begin with a review of the basic ideas on Poisson geometry in
Section 2 (see \cite{Vaisman} for a thorough study). Section 3
comprises some well-known facts about the reduction of a Poisson
manifold along with some new results and a somewhat original approach
to the subject. In section 4 we give a brief presentation of Poisson
Sigma models from the lagrangian point of view.

Section 5 is devoted to the study of general boundary conditions for
the Poisson-Sigma model. We show explicitly that coisotropy of the
brane is not essential and give the precise boundary conditions that
the fields must satisfy in the general case, obtaining the results of
\cite{CaFe3} as a particular case.

In Section 6 we carry out the hamiltonian study of the model for an
open string with our boundary conditions. It turns out that there
exist a Poisson and an anti-Poisson map from the phase space to the
branes at the endpoints of the string when in the latter ones the
(generalized) Dirac bracket obtained by Poisson reduction of $M$ is
considered.

Section 7 contains our conclusions as well as the discussion on 
the quantization of the model and future lines of research.

\section{Poisson geometry} \label{geometry}

Let $(\cal A,\cdot)$ be an associative, commutative algebra (we will
omit the symbol $\cdot$ for the commutative product on $\cal A$ in the
following) with unit over the real or complex numbers. $(\cal
A,\cdot,\{,\})$ is said to be a {\it Poisson algebra} if:

$(i)$ $\{,\}$ defines a Lie bracket on $\cal A$

$(ii)$ Leibniz rule (compatibility of both products) is satisfied, i.e.
$$\{x,yz\}=y\{x,z\}+\{x,y\}z,\ \forall x,y,z \in \cal A$$

When ${\cal A}$ is the algebra of smooth functions on a manifold 
$M$ the concept of  Poisson algebra is equivalent to that 
of {\it Poisson manifold}.
An $m$-dimensional Poisson manifold $(M,\Gamma)$ is a differentiable
manifold $M$ equipped with a bivector field $\Gamma$
that makes the algebra of smooth functions 
a Poisson algebra when the Poisson bracket of two
functions in $C^\infty(M)$ is given by the contraction of $\Gamma$:
$$\{f,g\}(p)=\iota(\Gamma_p) (df\wedge dg)_{p},\ p \in M$$ 
Taking local coordinates $X^i$ on $M$, $\Gamma^{ij}(X)=\{X^i,X^j\}$. The Jacobi
identity for the Poisson bracket reads in terms of $\Gamma^{ij}$:
$$
\Gamma^{ij}
\partial_{i}\Gamma^{kl}
+
\Gamma^{ik}
\partial_{i}\Gamma^{lj}
+
\Gamma^{il}
\partial_{i}\Gamma^{jk}
=0
$$
where summation over repeated indices is understood. 

\vspace{0.3 cm}

Define $\Gamma^\sharp:T^*M\rightarrow TM$ by
$$\beta(\Gamma^\sharp(\alpha))=\iota(\Gamma) (\alpha \wedge \beta),\ \alpha,\beta \in T^{*}M$$

By virtue of Jacobi identity, the image of $\Gamma^{\sharp}$,
$${\rm Im}(\Gamma^{\sharp}):=\bigcup_{p\in M}{\rm Im}(\Gamma^{\sharp}_{p})$$
is a completely integrable (general) differential distribution and $M$
admits a (generalized) foliation (see \cite{Vaisman} for definitions
of these concepts). $M$ is foliated into leaves which may have varying
dimensions. The Poisson structure can be consistently restricted to a
leaf and this restriction defines a non-degenerate Poisson structure
on it. That is why we will also refer to the leaves as {\it symplectic
leaves} and to the foliation as the {\it symplectic foliation} of $M$. This
result comes from a generalization of the classical Frobenius theorem
for regular distributions.

An example of a Poisson manifold is obtained by taking $M={\gl g}^{*}$,
where ${\gl g}$ is a Lie algebra. Hence, $M$ is a linear space and
the Poisson structure is the so called Kostant-Kirillov Poisson
structure that, for the linear functions, is given by
the Lie bracket of $\gl g$, i.e.
$$\{f,g\}=[f,g],\  f,g \in {\gl g}.$$

The symplectic leaves in this case correspond to the orbits under the
coadjoint representation of any connected Lie group $G$ with Lie
algebra $\gl g$ and have, in general, varying dimensions (in
particular, the origin is always a symplectic leaf).

Another way of defining a Poisson algebra for functions on $M$ is via
a presymplectic structure, i.e. a closed two-form $\omega\in\Lambda^2(M)$.
In this case the Poisson algebra ${\cal A}$ consists of functions
that possess a hamiltonian vector field, i.e. 
those functions $f\in{C}^\infty(M)$ for which the equation
$$\omega({\cal X},{\cal Y})={\cal Y}(f)$$ has a solution ${\cal X}\in{\gl X}(M)$ for any
${\cal Y}\in{\gl X}(M)$. Given $f_1,f_2\in{\cal A}$ with hamiltonian vector
fields ${\cal X}_1,{\cal X}_2$ respectively, $f_1f_2$ has the hamiltonian vector
field $f_1{\cal X}_2+f_2{\cal X}_1$ and then ${\cal A}$ is a subalgebra of
${C}^\infty(M)$ (if and only if $\omega$ is symplectic the
Poisson algebra induced by it gives $M$ the structure of a Poisson
manifold). The Poisson bracket is defined by
$$\{f_1,f_2\}=\omega({\cal X}_1,{\cal X}_2).$$ Note that in general the hamiltonian
vector field ${\cal X}_1$ for $f_1\in {\cal A}$ is not uniquely defined but the
ambiguities are in the kernel of $\omega$ and then it leads to a well
defined Poisson bracket.  Due to closedness of $\omega$,
$\{f_1,f_2\}\in{\cal A}$ and the Jacobi identitiy is satisfied. It is worth 
mentioning that in this case the center of ${\cal A}$ 
(Casimir functions) is the set of constant functions on $M$. 

Given two Poisson manifolds $(M_1,\Gamma_1)$, $(M_2,\Gamma_2)$ and a
differentiable map $F:M_1 \rightarrow M_2$, $F$ is a Poisson map if
$$\{f,g\}_2 \circ F = \{f\circ F,g\circ F\}_1,\ \forall f,g \in C^{\infty}(M_2)$$
and an anti-Poisson map if
$$\{f,g\}_2 \circ F = -\{f\circ F,g\circ F\}_1,\ \forall f,g \in C^{\infty}(M_2)$$

The concept of Poisson map can be extended to the algebraic setup.
Given two Poisson algebras $({\cal A}_1,\{.,.\}_1)$, and  $({\cal A}_2,\{.,.\}_2)$
and a homomorphism of (abelian, associative) algebras, 
$\Phi:{\cal A}_2\rightarrow{\cal A}_1$ we say that $\Phi$ is (anti-)Poisson if it is 
also a 
(anti-)homomorphism of Poisson algebras, i.e. 
$$\Phi(\{f,g\}_2)=(-)\{\Phi(f),\Phi(g)\}_1.$$
In this paper we consider the case in which
the Poisson algebras are subalgebras of
the space of functions on certain manifolds
and the homomorphism of algebras is induced by a map
between the manifolds themselves.  

\section{Reduction of Poisson manifolds} \label{reduction}

Let $C$ be a closed submanifold of $(M,\Gamma)$. Can we define in a
natural way a Poisson structure on $C$? The answer is negative, in
general. What we can always achieve is to endow a certain subset of
$C^{\infty}(C)$ with a Poisson algebra structure. The canonical
procedure below follows in spirit reference \cite{Kimura}, although we
present some additional, new results.

We adopt the notation ${\cal A} = C^{\infty}(M)$ and take the ideal 
(with respect to the pointwise product of functions in ${\cal A}$. 
We will use the term Poisson ideal when we refer to an ideal 
with respect to the Poisson bracket).
$${\cal I} =\{f \in {\cal A} \vert f(p) = 0,\ p \in C\}$$

Define ${\cal F}\subset{\cal A}$ as the
set of {\it first-class functions}, also called the {\it normalizer} of ${\cal I}$,
$${\cal F}=\{f\in{\cal A} \vert \{f,{\cal I}\}\subset{\cal I}\}.$$ Note that due to the
Jacobi identity and the Leibniz rule ${\cal F}$ is a Poisson subalgebra of
${\cal A}$ and ${\cal F}\cap{\cal I}$ is a Poisson ideal of ${\cal F}$. Then, we have
canonically defined a Poisson bracket in the quotient
${\cal F}/({\cal F}\cap{\cal I})$.  However, this is not what we want, as our problem
was to find a Poisson bracket in $C^{\infty}(C)\cong {\cal A}/{\cal
I}$ (or, at least, in a subset of it). To that end we define an
injective map
\begin{eqnarray}\label{mapphi}
 \begin{matrix}\phi:&{\cal F}/({\cal F}\cap{\cal I})&\longrightarrow&{\cal A}/{\cal I}\cr
&f+{\cal F}\cap{\cal I}&\longmapsto&f+{\cal I}
\end{matrix}
\end{eqnarray}
$\phi$ is an homomorpism of abelian, associative algebras with unit 
and then induces a Poisson algebra structure $\{.,.\}_C$ in the 
image, that will be denoted by ${\cal C}(\Gamma,M,C)$, 
i.e.:
\begin{eqnarray}\label{ourDirac}
\{ f_1+{\cal I}, f_2+{\cal I}\}_{_C}=
\{ f_1, f_2\}+{\cal I}.\qquad f_1,f_2\in {\cal F}.
\end{eqnarray}

\vspace{0.3 cm}

{\it Remarks:} 
\begin{itemize}
\item{\it Poisson reduction} is a generalization of the 
symplectic reduction in the following sense:\hfill\break
If the original Poisson structure is non-degenerate, it induces 
a symplectic structure $\omega$ in $M$. 
Then, we may canonically define on $C$ the
closed two-form $i^*\omega$,
where $i:C\rightarrow M$ is the inclusion map.
As described before, this presymplectic two-form in $C$ 
defines a Poisson algebra for a certain subset of  
${C}^\infty(C)\cong {\cal A}/{\cal I}$. The Poisson algebra obtained 
this way  coincides with the one defined above.

\item Note that the elements of ${\cal F}\cap{\cal I}$ are, in the language of
physicists, the generators of {\it gauge transformations} or, in Dirac's
terminology, the {\it first-class constraints}.
\end{itemize}

\vspace{0.3 cm}

The problem is that in general $\phi$ is not onto and $C$
cannot be made a Poisson manifold. The goal now is to use the
geometric data of the original Poisson structure to interpret the
algebraic obstructions.

Let $N^*C$ (or ${\rm Ann}(TC)$) be the conormal bundle of $C$ (or
annihilator of $TC$) i.e., the subbundle of the pull-back
$i^{*}(T^{*}M)$ consisting of covectors that kill all vectors in $TC$.
Now one has the following

\vspace{0.3 cm}

{\bf Theorem 1:}

{\it\parindent 0pt 
Assume that:

a) ${\rm dim}(\Gamma^\sharp_p(N_p^*C)+ T_pC)= k,\ \forall p \in C$, and

b) $\Gamma^\sharp_p(N_p^*C)\cap T_pC=\{0\},\ \forall p \in C$

Then the map $\phi$ of (\ref{mapphi}) is an isomorphism of associative, 
commutative algebras with unit.
}
\vspace{0.3 cm}

{\it Proof:} Condition b) implies that
$$T^*_pM = {\rm Ann}(\Gamma^\sharp_p(N_p^*C)\cap
T_pC)=N_p^*C+\Gamma^{\sharp-1}_p(T_pC)$$
and then, 
$\Gamma_p^{\sharp}(T^*_pM)\subseteq \Gamma^\sharp_p(N_p^*C)+ T_pC, \quad
\forall p \in C$.

Now define a smooth bundle map:
$$\Upsilon:N^*C+ TC\longrightarrow i^*TM$$
that maps $(\alpha_p,v_p)\in N_p^*C+T_pC$ to 
$\Gamma^\sharp_p \alpha_p + v_p$. Due to condition 
a) the map is of constant rank and then every 
smooth section of its image has a smooth preimage.

Take $f\in {\cal A}$.  As shown before $\Gamma^\sharp_p (df)_p\in
\Gamma^\sharp_p(N_p^*C)+ T_pC$ for any $p\in C$. Then, the restriction to
$C$ of $\Gamma^\sharp df$ is a smooth section of the image of
$\Upsilon$. Let $(\alpha, v)$ be a smooth section of $N^*C+TC$ with
$\Upsilon(\alpha, v)_p=\Gamma_p^\sharp (df)_p$ for $p \in C$. Now for any
section $\alpha$ of $N^*C$ there exists a function $g\in{\cal I}$ such that
$\alpha_p = (dg)_p$ for any $p\in C$.

Hence, one has that $\tilde f=f-g\in {\cal F}$ and 
$\phi(\tilde f+ {\cal F}\cap{\cal I})= f+ {\cal I}$.
$\,$\hfill$\Box$\break

\vspace{0.5 cm}
When ${\rm dim}(\Gamma^\sharp_p(N_p^*C)+ T_pC)={\rm dim}(M)$ we can
choose locally a basis $\{g_n\}$ of regular second-class
constraints. The matrix of the Poisson brackets of the
constraints $G_{mn}=\{g_m,g_n\}$ is invertible on $C$ and
the Poisson bracket of (\ref{ourDirac}) is:
\begin{eqnarray}\label{Dirac}
\{f +{\cal I},f'+{\cal I}\}_{_C}=\{f,f'\}-\sum_{m,n=1}^\nu\{f,g_n\}G^{-1}_{nm}
\{g_m,f'\}+{\cal I}
\end{eqnarray}
which is the usual definition of the Dirac bracket restricted to $C$. 
In this case, of course, every function on $C$ has a well-defined 
Poisson bracket and we get a  Poisson structure on $C$. 
\vspace{0.2cm}

Condition a) of Theorem 1 is not necessary as can be shown in the following

{\bf Example 1:}

Take $M={\gl sl}(2)^*$. In coordinates $(x_1,x_2,x_3)$ the linear
Poisson bracket is given by $\{x_i,x_j\}=\epsilon^{ijk}x_k$. Now
define $C$ by the constraints: $x_1=0, x_2=0$. Clearly,
$${\rm dim}(\Gamma^\sharp_p(N_p^*C)+ T_pC)=
\begin{cases}
3\quad {\rm for}\ p \not= 0\cr
1\quad {\rm for}\ p = 0
\end{cases} $$
and for any $f\in {C}^\infty(M)$ we may define $\tilde
f=f-x_1\partial_1 f -x_2\partial_2 f\in{\cal F}$ such that $\phi(\tilde
f+{\cal F}\cap{\cal I})=f+{\cal I}$, i.e. $\phi$ is onto. The Poisson structure
induced in this case is, of course, zero.
\vspace{0.5 cm}

Condition b), however, is indeed necessary:
\vspace{0.3 cm}

{\bf Theorem 2:}

{\it \noindent 
If map $\phi$  of (\ref{mapphi})
is onto then  $\Gamma^\sharp_p(N_p^*C)\cap T_pC=\{0\}$}

\vspace{0.3 cm}

{\it Proof:} Assume that $\exists v_p\not=0, 
v_p \in \Gamma^\sharp_p(N_p^*C)\cap T_pC$. 
It is enough to take a function $f\in{\cal A}$ such that 
its directional derivative at $p$ in direction $v_p$ does not
vanish. Then $f+{\cal I}$ is not in the image of $\phi$.
$\,$\hfill$\Box$\break

\vspace{0.5 cm}

This result tells us that when $\Gamma^\sharp_p(N_p^*C)\cap
T_pC\not=\{0\}$ one cannot endow $C$ with a Poisson structure. The only
functions on $C$ that have got a well-defined Poisson bracket
(i.e. the physical observables) are those in the image of $\phi$. 
On the other hand, it is easy to see that all functions in the image of $\phi$
belong to the subalgebra of gauge invariant functions
$${\cal A}_{inv}=\{ f\in{\cal A} | \{f,{\cal F}\cap{\cal I}\}\subset{\cal I}\}.$$
One may wonder when the physical observables are precisely
the gauge invariant functions.
A sufficent condition is given by the following

\vspace{0.3 cm}

{\bf Theorem 3:}

{\it \parindent 0pt If 
${\rm dim}(\Gamma^\sharp_p(N_p^*C)+ T_pC)= k,\ \forall  p\in C$,
 then $\phi ({{\cal F}}/{{\cal F}}\cap{{\cal I}})={\cal A}_{inv}/{\cal I}$.}

\vspace{0.3 cm}

Before proving the theorem we will establish a Lemma
that will be useful in the following.

\vspace{0.3 cm}

{\bf Lemma 1:}

{\it \parindent 0pt
The following two statements are equivalent:

a) ${\rm dim}(\Gamma^\sharp_p(N_p^*C)+ T_pC)=k,\ \forall p \in C$

b) $\Gamma_p^{\sharp-1}(T_pC)\cap N^*C=\{(dg)_p | g\in{\cal F}\cap{\cal I}\},\
\forall p \in C$.
}
\vspace{0.3 cm}

{\it Proof:}

a) $\Rightarrow$ b): Assume that ${\rm dim}(\Gamma^\sharp_p(N_p^*C)+ T_pC)$ is
constant on $C$. Then, ${\rm Ann}_p(\Gamma^\sharp(N^*C)+
TC)=\Gamma_p^{\sharp-1}(T_pC)\cap N_p^*C$ is also of constant dimension
and $\Gamma^{\sharp-1}(TC)\cap N^*C$ is a subbundle of $N^*C$ whose
fiber at every point of the base is spanned by a set of sections. For
every section $\alpha$ of this subbundle there exists $g\in{\cal I}$ such
that $\alpha_p=(dg)_p$.  But since $(dg)_p\in
\Gamma_p^{\sharp-1}(T_pC)$, it follows that $g\in{\cal F}\cap{\cal I}$.

The other inclusion is trivial as differential of first-class
constraints are in $N^*C$ (because they are constraints) and their
hamiltonian vector fields transform constraints into constraints
(because they are first-class) so their restrictions to $C$ are in $TC$.

b) $\Rightarrow$ a): Assuming b) one has that ${\rm
dim}(\Gamma_p^{\sharp-1}(T_pC)\cap N_p^*C)$ is a lower semicontinuous
function on $C$ 
because the fiber of
$\Gamma_p^{\sharp-1}(T_pC)\cap N_p^*C$ at every point is spanned by local
sections (see ref. \cite{Vaisman}). For the same reason, ${\rm dim}(\Gamma^\sharp_p(N_p^*C)+ T_pC)$
is also lower semicontinuous. But from the relation
$\Gamma^\sharp_p(N_p^*C)+ T_pC= {\rm Ann}_p(\Gamma^{\sharp-1}(TC)\cap N^*C)$
we infer that ${\rm dim}(\Gamma^\sharp_p(N_p^*C)+ T_pC)$ is upper
semicontinuous,
so it is continuous and, being integer valued it is indeed constant.
$\,$\hfill$\Box$\break

\vspace{0.5 cm}

{\it Proof of Theorem 3:} First note that $f\in{\cal A}_{inv}$ implies that
$\Gamma_p^\sharp(df)_p\in {\rm Ann}_p(\{dg| g\in{\cal F}\cap{\cal I}\})$.  But from
the previous Lemma we have that the latter is equal to
$\Gamma^\sharp_p(N_p^*C)+ T_pC$.

Then, $\forall f\in{\cal A}_{inv}$ 
one has $\Gamma_p^\sharp(df)_p\in \Gamma^\sharp_p(N_p^*C)+ T_pC$.
And from here on the proof is like that of Theorem 1.
$\,$\hfill$\Box$\break

\vspace{0.5 cm}

At first sight we might expect a result analogous to Theorem 2 for the
case with gauge transformations in the constrained submanifold, namely
that a necessary condition for $\phi$ mapping onto the space of gauge
invariant functions on $C$ is that the space of hamiltonian vector
fields of first-class constraints at every point coincides with
$T_pC\cap \Gamma_p^\sharp(N^*C)$. This is not true, however, as shown
by the following example in which the spaces above differ in some
points whereas the image of map $\phi$ of (\ref{mapphi}) is
${\cal A}_{inv}/{\cal I}$.

{\bf Example 2:}

Take $M={\mathbb R}^6=\{(x_1,x_2,x_3,p_1,p_2,p_3)\}$ with the standard
Poisson bracket $\{p_i,x_j\}=\delta_{ij}$. Now consider the
constraints 
$$g_i:=p_i-x_ix_{\sigma(i)}\quad i=1,2,3$$ 
with $\sigma$ the cyclic permutation 
of $\{1,2,3\}$ s. t. $\sigma(1)=2$.
In this case
$${\rm dim}(\Gamma^\sharp_m(N_m^*C)\cap T_mC)=
\begin{cases}1\quad {\rm for}\ m \not= 0\cr
3\quad {\rm for}\ m = 0
\end{cases}
$$
while the gauge transformations 
are restrictions to $C$ of hamiltonian vector fields  
of $fg$ with $f\in{C}^\infty(M)$ and 
$g=x_2g_1+x_3g_2+x_1g_3$.
It implies that at $m=0$ the gauge transformations 
vanish and, hence, they do not fill
$\Gamma^\sharp_m(N_m^*C)\cap T_mC$.

We will show that the image of map $\phi$ of (\ref{mapphi})
is ${\cal A}_{inv}/{\cal I}$. 
In every class  of ${\cal A}_{inv}/{\cal I}$ we may take the only 
representative independent of the $p_i$'s. 
Gauge invariant functions 
$f(x_1,x_2,x_3)$ are then characterized by:
$$(x_2\partial_{x_1}+x_3\partial_{x_2}+x_1\partial_{x_3})f=0,$$
and for any of them we may define
$$\tilde f=f+\sum_i a_i g_i$$
with $a_i$ smooth, given by
\begin{eqnarray}
a_1(x_1,x_2,x_3)&=&{1\over x_1}
[\partial_{x_2}f(x_1,x_2,x_3)-\partial_{x_2}f(0,x_2,x_3)]\cr
a_2(x_1,x_2,x_3)&=&{1\over x_1}
[\partial_{x_1}f(x_1,x_2,x_3)-\partial_{x_1}f(0,x_2,x_3)]\cr
a_3(x_1,x_2,x_3)&=&
{1\over x_2}
\partial_{x_2}f(0,x_2,x_3)= -{1\over x_3}
\partial_{x_1}f(0,x_2,x_3)
\end{eqnarray}
Now $\tilde f$ is first class and 
$\phi(\tilde f+{\cal F}\cap{\cal I})=f+{\cal I}.$
This shows that in this case the image of $\phi$ fills ${\cal A}_{inv}/{\cal I}$.
\vspace{0.5 cm}

In general, if ${\rm dim}(\Gamma^\sharp_p(N_p^*C)+ T_pC)$ is
not constant on $C$ we cannot define a Poisson bracket even in the set
of gauge invariant functions. The only thing we can
assert is that we have a Poisson algebra on the subset 
of $C^{\infty}(C)$ given by the image of $\phi$.
However, an efficient description of the functions in the image 
(the space of {\it observables}) is not available in the general case. 
\vspace{0.5 cm}

{\it Remark:}

$C$ is said coisotropic if $\Gamma^{\sharp}(N^{*}C) \subseteq TC$. For
such a $C$, ${\cal I} \subseteq {\cal F}$. Then, ${\cal F} \cap {\cal I} = {\cal I}$, ${\cal F} =
{\cal A}_{inv}$ and $\phi$ is the identity map.

\subsection{Poisson-Dirac submanifolds}

In this subsection we would like to make contact between the results
and terminology of this section in absence of gauge transformations
and those appearing in two papers by Crainic and Fernandes
\cite{CraFer} and Vaisman \cite{Vai02}.

If $\Gamma^\sharp_p(N_p^*C)\cap T_pC=\{0\},\ \forall p \in C$, $C$ is
called {\it pointwise Poisson-Dirac} in \cite{CraFer}. If, in
addition, the induced Poisson bivector defined therein is smooth $C$
is said a {\it Poisson-Dirac submanifold}. It is clear that $\phi$
onto implies that $C$ is a Poisson-Dirac submanifold. The following is
an example in which $\phi$ is not onto while $C$ is still a
Poisson-Dirac submanifold, being possible to endow it with a Poisson
structure.

{\bf Example 3:}

Consider $M={\mathbb R}^4=\{(x_1,x_2,p_1,p_2)\}$ with Poisson structure
$\{p_i,x_j\}=\delta_{ij}x_{i}{\rm exp}(-1/x_{i}^{2})$ smoothly
extended to $x_i=0$ and $C$ defined by the constraints $g_1 = p_1 -
{x_2^{2}}/{2}$, $g_2 = p_2 + {x_1^{2}}/{2}$. We can take $\sigma_i :=
x_i$ as coordinates on $C$.

$\Gamma^\sharp_p(N_p^*C)\cap T_pC=\{0\}$ on $C$ but $\phi$ is not
onto. For instance, take $f_i = x_i \in C^{\infty}(M)$. If we try to
find a first-class function in the class $f_1 + {\cal I}$ (its pre-image by
$\phi$) we obtain for $x_i\neq 0$
$$\tilde{f_1}:= f_1-\frac{x_1{\rm exp}(-1/x_{1}^{2})}{x_1^{2}{\rm exp}(-1/x_{1}^{2})+x_2^{2}{\rm exp}(-1/x_{2}^{2})}$$
which fails to extend continuously to $x_i = 0$. Then, $f_1$ does not belong to the image of $\phi$. However, the hamiltonian vector field associated to this singular $\tilde{f_1}$ is smooth and we can define a Poisson structure on $C$:
$$\Gamma_{_C}^{12}(\sigma_1,\sigma_2) = \{\tilde{f_1},f_2\}(\sigma_1,\sigma_2,{1\over2}\sigma_2^2,-{1\over2}\sigma_1^2) = \frac{\sigma_1\sigma_2}{\sigma_1^2{\rm exp}(1/{\sigma_2^2})+\sigma_2^2{\rm exp}(1/{\sigma_1^2})}$$

\vspace{0.5 cm}

If $\phi$ is onto and, in addition, ${\rm
dim}(\Gamma^\sharp_p(N_p^*C)+ T_pC)$ is constant on $C$ (i.e. the
situation of Theorem 1), we have what is called in \cite{CraFer} a
{\it constant rank Poisson-Dirac submanifold}.

\vspace{0.5 cm}

Following \cite{Vai02}, define a {\it normalization} of $C$ by a
{\it normal bundle} $\nu C$ as a splitting $TM\vert_{C}=TC\oplus\nu C$. For
every $p\in C$ there exists a neighborhood $U$ where we can choose
adapted coordinates $(g^A,y^{\alpha})$ such that, locally,
$g^A\vert_{C\cap U}=0$ and $y^{\alpha}$ are coordinates on
$C\cap U$. Vaisman calls $\nu C$ {\it algebraically} $\Gamma${\it
-compatible} if, in these coordinates,
$\Gamma^{A \alpha}\vert_{C}=0$. The relation with our map $\phi$ is given by the following
\vspace{0.5 cm}

{\bf Theorem 4:} 

\noindent $\phi$ {\it is onto iff there exists an algebraically} $\Gamma${\it-compatible normal bundle.}
\vspace{0.3 cm}

{\it Proof:}

Only local properties in a neighborhood of each point of $C$ matter
for this proof.

$\Rightarrow )$ Let $(g^A,z^\alpha)$ be local coordinates such that
$C$ is locally defined by $g^A=0$ and $z^\alpha$ are coordinates on
$C$. Take the pre-image by $\phi$ of the coordinate functions
$z^\alpha$ and denote them by $y^\alpha$. $(g^A,y^\alpha)$ are local
coordinates such that $\Gamma^{A\alpha}\vert_{C}=0$.

$\Leftarrow )$ For any $f(g^A,y^\alpha) \in C^\infty(U)$,
$\tilde{f}(g^A,y^\alpha)=f(0,y^\alpha)\in {\cal F}$ and $\tilde{f}-f \in
{\cal I}$. Then, $\phi$ is onto.$\,$\hfill$\Box$\break

\section{Poisson-Sigma models} \label{PSmodels}
 
The Poisson-Sigma model is a two-dimensional topological Sigma model
defined on a surface $\Sigma$ and with a finite dimensional Poisson
manifold $(M,\Gamma)$ as target.

The fields of the model are given by a bundle map $(X,\psi): T\Sigma
\rightarrow T^{*}M$ consisting of a base map $X:\Sigma \rightarrow M$
and a 1-form $\psi$ on $\Sigma$ with values in the pullback by $X$ of
the cotangent bundle of $M$.  The action functional has the form
\begin{eqnarray} \label{PS}
S_{P\sigma}(X,\psi)=\int_\Sigma \langle dX,\wedge\psi\rangle - 
{1\over2}\langle\Gamma\circ X,\psi\wedge\psi\rangle
\end{eqnarray}
where $\langle,\rangle$ denotes the pairing between vectors and
covectors of $M$.

If $X^{i}$ are local coordinates in $M$, $\sigma
^{\mu},\ \mu=1,2$ local coordinates in $\Sigma$, $\Gamma^{ij}$ the
components of the Poisson structure in these coordinates and
$\psi_{i}=\psi_{i\mu}d{\sigma}^{\mu}$, the action reads
\begin{eqnarray} \label{PScoor}
S_{P\sigma}(X,\psi)=\int_\Sigma dX^{i}\wedge \psi_{i}-
{1\over2}\Gamma^{ij}(X)\psi_{i}\wedge \psi_{j}
\end{eqnarray}

It is straightforward to work out the equations of motion in the bulk:
\begin{subequations}\label{eom}
\begin{align}
&dX^{i}+\Gamma^{ij}(X)\psi_{j}=0 \label{eoma} \\ 
&d\psi_{i}+{1\over2}\partial_{i}\Gamma^{jk}(X)\psi_{j}\wedge \psi_{k}=0 \label{eomb}
\end{align}
\end{subequations}

One can show (\cite{BojoStrobl}) that for solutions
of \eqref{eoma} the image of $X$ lies within 
one of the symplectic leaves  of the foliation of $M$.

Under the infinitesimal transformation
\begin{subequations}\label{symmetry}
\begin{align}
&\delta_{\epsilon}X^{i}=
\Gamma^{ji}(X)\epsilon_{j}\label{symmetrya}\\
&\delta_{\epsilon}\psi_{i}=d\epsilon_{i}+\partial_{i}\Gamma^{jk}(X)\psi_{j}
\epsilon_{k}\label{symmetryb}
\end{align}
\end{subequations}
where $\epsilon=\epsilon_{i}dX^i$ is a section of $X^*(T^*(M))$,
the action (\ref{PScoor}) transforms by a boundary term
\begin{eqnarray} \label{symmS}
\delta_{\epsilon}S_{P\sigma}=-\int_\Sigma d(dX^i \epsilon_i).
\end{eqnarray}

{}Formula (\ref{symmetry}) is not the most general transformation that
leaves the action invariant up to a boundary term, but it gives a 
complete set of gauge
transformations in the sense that any symmetry of the action is of
type (\ref{symmetry}) up to terms vanishing on-shell (the so-called
{\it trivial gauge transformations} of \cite{TeHe}). Then, it is not
surprising that the commutator of two consecutive gauge tansformations
of type (\ref{symmetry}) is not of the same form, i.e.
\begin{subequations}\label{commg}
\begin{align}
 [\delta_\epsilon,\delta_{\epsilon'}]X^i&=\delta_{[\epsilon,\epsilon']^*} X^i \label{commga} \\
[\delta_\epsilon,\delta_{\epsilon'}]\psi_i&=
\delta_{[\epsilon,\epsilon']^*} \psi_i
+\epsilon_k\epsilon_{l}'\partial_i\partial_j 
\Gamma^{kl}(dX^{j}+\Gamma^{js}(X)\psi_{s})\label{commgb}
\end{align}
\end{subequations}
where $[\epsilon,\epsilon']^{*}_k :=
\partial_k\Gamma^{ij}(X)
\epsilon_i\epsilon'_j$.
 Note that the term in parenthesis in
\eqref{commgb} is the equation of motion \eqref{eoma} and then, 
as expected, it vanishes on-shell.  

In this section we have analysed the equations of motion
and gauge invariance in the bulk. In the following one we will address 
the subject of boundary conditions for the fields
and how they affect the gauge transformations.

\section{Boundary conditions} \label{BC}

We study now the previous model on a surface with boundary and search
for the boundary conditions (BC) which make the theory consistent.

In order to preserve the topological character of the theory one must
choose the BC independent of the point of the boundary, as far as we
move along one of its connected components.  For the sake of clarity we
will restrict ourselves in this section to one connected component
(without mentioning it explicitly). In the next section we will
discuss the relation between the BC in the possible different
connected components of the boundary.

In surfaces with boundary a new term appears in the variation of the
action under a change of $X$ when performing the integration by parts:
\begin{eqnarray}
\delta_{X} S=\int_{\partial\Sigma}\delta X^i\psi_i-
\int_\Sigma\delta X^i(d\psi_{i}+
{1\over2}\partial_{i}\Gamma^{jk}(X)\psi_{j}\wedge \psi_{k})
\end{eqnarray}
 
The BC must cancel the surface term.

Let us take the field
\begin{eqnarray}\label{BCX}
X|_{\partial \Sigma}:\partial \Sigma\rightarrow C
\end{eqnarray}
for an arbitrary (for the moment) closed embedded submanifold $C$ of $M$
(brane, in a more stringy language). Then $\delta X\in T_{_X} C$ at
every point of the boundary and the contraction of $\psi=\psi_{i}dX^i$
with vectors tangent to the boundary (that we will denote by
$\psi_t=\psi_{it}dX^i$) must belong to $N^*_{_X}(C)$ (the fiber over $X$
of the conormal bundle of $C$).

On the other hand, by continuity, the equations of motion in the bulk
must be satisfied also at the boundary. In particular,
$$\partial_t X=\Gamma^\sharp\psi_t$$
where by $\partial_t$ we denote the
derivative along the vector on $\Sigma$ tangent to the boundary.  As
$\partial_t X$ belongs to $T_{_X}C$ it follows that
$\psi_t\in\Gamma^{\sharp-1}_{_X}(T_{_X}(C))$.

Both conditions for $\psi_t$ imply that
\begin{eqnarray}\label{BCpsi}
\psi_t(m)\in\Gamma^{\sharp-1}_{_{X(m)}}(T^{}_{_{X(m)}}C)\cap N_{_{X(m)}}^*C,
\mbox{ for any }m\in \partial\Sigma 
\end{eqnarray}
which is the boundary condition we shall take for $\psi_t$.

We should check now that the BC are consistent with
the gauge transformations (\ref{symmetry}).

In order to cancel the boundary term (\ref{symmS})
$\epsilon\vert_{\partial\Sigma}$ must be a smooth section of
$N^{*}(C)$ and if (\ref{symmetry}) is to preserve the boundary
condition of $X$, $\epsilon\vert_{\partial\Sigma}$ must belong to
$\Gamma^{\sharp-1}(TC)$. Hence,
\begin{eqnarray}\label{BCepsilon}
\epsilon(m)\in\Gamma^{\sharp-1}_{_{X(m)}}(T^{}_{_{X(m)}}C)\cap N_{_{X(m)}}^*C,
\mbox{ for any }m\in \partial\Sigma 
\end{eqnarray}

Next, we shall show that the the gauge transformations
(\ref{symmetry}) with (\ref{BCepsilon}) also preserve
(\ref{BCpsi}). At this point we must restrict ourselves to the case in
which
\begin{equation}\label{consrank}
{\rm dim}(\Gamma^\sharp_p(N_p^*C)+ T_pC)=k, \mbox{ for any } p\in C
\end{equation}
In this case we can choose, at least locally, a set of regular
constraints with a maximal number (${\rm dim}(M)-{\rm
dim}(\Gamma^\sharp_p(N_p^*C)+ T_pC)$) of first-class ones. Let
$\{\chi^a\}$ be the set of first-class constraints and $\{\gamma^A\}$
that of second-class ones.  Local regularity means that for every point in
$C$ there is a neighborhood $U\subset C$ and a choice of constraints,
such that differentials of the constraints at $p\in U$ span
$N^*_p(C)$. $U$ can be chosen so that we can also find coordinates
$\{y^\alpha\}$ on $C$.  Then $(y^\alpha,\chi^a,\gamma^A)$ form
a set of local coordinates for an open subset of $M$ containing $U$.

In these coordinates the Poisson structure satisfies:
\begin{eqnarray} \label{localcoor}
\Gamma^{ab}\vert_C = 0,\quad \Gamma^{aA}\vert_C = 0,\quad {\rm det}(\Gamma^{AB})\vert_C \neq 0
\end{eqnarray}

The boundary condition (\ref{BCpsi}) translates in these coordinates
into $\psi_t = \psi_{at}d\chi^a$. Hence, we must show that
$\delta\psi_{\alpha t}=\delta\psi_{At} = 0$. Recalling (\ref{BCepsilon}) we
also may write $\epsilon\vert_C = \epsilon_a d\chi^a$ and therefore,
$$\delta\psi_{\alpha t}=
\partial_\alpha\Gamma^{ab}\vert_C\psi_{at}\epsilon_b\vert_C$$
which vanishes because $\Gamma^{ab}\vert_C = 0 \Rightarrow
\partial_\alpha\Gamma^{ab}\vert_C = 0$.

Showing that
$$\delta\psi_{At}=\partial_A\Gamma^{ab}\vert_C\psi_{at}\epsilon_b\vert_C$$
also vanishes on $C$ is more tricky, but it does, as a consequence of
the Jacobi identity:
\begin{eqnarray}
\Gamma^{AB} 
\partial_A\Gamma^{ab}
+
\Gamma^{\alpha B} 
\partial_\alpha\Gamma^{ab}
+
\Gamma^{cB} 
\partial_c\Gamma^{ab}
&&\cr
+ \Gamma^{Ab} 
\partial_A\Gamma^{Ba}
+
\Gamma^{\alpha B} 
\partial_\alpha\Gamma^{Ba}
+
\Gamma^{cb} 
\partial_c\Gamma^{Ba}
&&\cr
+ \Gamma^{Aa} 
\partial_A\Gamma^{bB}
+
\Gamma^{\alpha a} 
\partial_\alpha\Gamma^{bB}
+
\Gamma^{ca} 
\partial_c\Gamma^{bB}
&=& 0
\end{eqnarray}

Evaluating on $C$ and using $\Gamma^{ab}\vert_C = 
\Gamma^{aA}\vert_C=0$ and $\partial_\alpha\Gamma^{ab}\vert_C = 
\partial_\alpha\Gamma^{aA}\vert_C=0$,
one may check that all terms except the first one vanish. Then,
$$\Gamma^{AB}\vert_C 
\partial_A\Gamma^{ab}\vert_C
= 0$$
Using now that $\Gamma^{AB}\vert_C$ is invertible, we conclude that
$\partial_A\Gamma^{ab}\vert_C = 0$ and then $\delta\psi_{At} = 0$.
A similar derivation proves that the gauge
transformations close on-shell at the boundary (see (\ref{commg})). 

At this point one might want to weaken somehow the condition
(\ref{consrank}) to allow for more general BC. Firstly, we notice
that some restriction must be imposed, as the existence of a maximal
number of regular first-class constraints seems to be
essential. Recall Example 2 of section 3 for a case in which gauge
transformations do not preserve the BC of
$\psi_t$. In this case the first class constraints with non-zero
differential on $C$ are generated by $x_2g_1+x_3g_2+x_1g_3$ which is
not regular at $0$.

A possible generalization of condition (\ref{consrank}) is to assume
that ${\rm dim}\{ (dg)_p | g\in{\cal F}\cap{\cal I}\}$ is constant on $C$ . With
this assumption we may choose a maximal number of regular first-class
constraints and the previous choice of coordinates works.  In this
case, however, ${\rm det}(\Gamma^{AB})\vert_C$ might be zero at some
points, but only in the complement of an open dense set. An argument
of continuity shows then that $\delta\psi_{At} = 0$ everywhere.

\section{Hamiltonian analysis of the Poisson-Sigma model}

We proceed to the hamiltonian study of the model with the BC of the
previous section (in each connected component of the boundary) when
$\Sigma = [0,\pi]\times {\mathbb R}$ (open string). The fields in the
hamiltonian formalism are a smooth map $X:[0,\pi]\rightarrow M$ and a
1-form $\psi$ on $[0,\pi]$ with values in the pull-back $X^* T^*(M)$;
in coordinates, $\psi=\psi_{i\sigma} dX^i d\sigma$.

Consider the infinite dimensional manifold of smooth maps $(X,\psi)$
with canonical symplectic structure $\Omega$. The action of $\Omega$
on two vector fields (denoted for shortness $\delta,\delta'$) reads
\begin{eqnarray} \label{Omega}
\Omega(\delta,\delta') = \int_{0}^{\pi}(\delta X^i \delta' 
\psi_{i\sigma} - \delta' X^i \delta \psi_{i\sigma})d\sigma
\end{eqnarray}

The phase space $P(M;C_0,C_{\pi})$ of
the theory is defined by the constraint:
\begin{eqnarray} \label{constraint}
\partial_{\sigma}X^i + \Gamma^{ij}(X)\psi_{j\sigma} = 0
\end{eqnarray}
and BC $X(0)\in C_0$ and $X(\pi)\in C_\pi$ for
two closed submanifolds $C_{u}\subset M$, ${u}=0,\pi$.

This geometry, with a boundary consisting of two connected components,
raises the question of the relation between the BC at both ends. Note
that due to eq. (\ref{constraint}) $X$ varies in $[0,\pi]$ inside a
symplectic leaf of $M$. This implies that in order to have solutions
the symplectic leaf must have non-empty intersection both with $C_0$
and $C_{\pi}$.  In other words, only points of $C_0$ and $C_{\pi}$
that belong to the same symplectic leaf lead to points of
$P(M;C_0,C_{\pi})$. In the following we will assume that this
condition is met for every point of $C_0$ and $C_\pi$ and
correspondingly for the tangent spaces. That is, if we denote by $J_0,
J_\pi$ the maps
\begin{eqnarray}
J_0:P(M,C_0,C_\pi)&\longrightarrow & C_0\cr
(X,\psi)&\longmapsto& X(0).
\end{eqnarray}
and 
\begin{eqnarray}
J_\pi:P(M,C_0,C_\pi)&\longrightarrow & C_\pi\cr
(X,\psi)&\longmapsto& X(\pi).
\end{eqnarray}
we assume that both maps are surjective submersions.

Vector fields tangent to the phase space satisfy the linearization of
(\ref{constraint}), i.e. $\delta\psi_{j\sigma}$ and $\delta X^i$ 
are such that
\begin{eqnarray} \label{linearization}
\partial_{\sigma}\delta X^i = \partial_j\Gamma^{ki}(X)
\psi_{k\sigma}\delta X^j + \Gamma^{ji}\delta \psi_{j\sigma}
\end{eqnarray}
with $\delta X(u)\in T_{_{X(u)}}C_u$, $u=0,\pi$.

The solution to the differential equation (\ref{linearization}) is
(\cite{CaFe4})
\begin{eqnarray} \label{solution}
\delta X^{i}(\sigma)= R^{i}_{j}(\sigma,0)\delta X^{j}(0) - 
\int_{0}^{\sigma}R^{i}_{j}(\sigma,\sigma')\Gamma^{jk}(X(\sigma'))
\delta \psi_{k\sigma}(\sigma')d\sigma'
\end{eqnarray}
where $R$ is given by the path-ordered integral
$$
R(\sigma,\sigma')
= \overleftarrow{P {\rm exp}
[\int_{\sigma'}^{\sigma}A_{\sigma}(z)dz]},
\qquad A^{i}_{j}(z)=(\partial_j\Gamma^{ki})(X(z))\psi_{k\sigma}(z).$$

The canonical symplectic 2-form is only presymplectic when restricted
to $P(M;C_0,C_{\pi})$. The kernel is given by:
\begin{eqnarray} \label{Hamsymm}
&&\delta_{\epsilon}X^{i}=\epsilon_{j}\Gamma^{ji}(X)\cr
&&\delta_{\epsilon}\psi_{i}=d\epsilon_{i}+\partial_{i}\Gamma^{jk}(X)\psi_{j}
\epsilon_{k}
\end{eqnarray}
where $\epsilon$, a section of $X^*(T^*M)$, is subject to the BC
$$\epsilon({u})\in \Gamma_{_{X({u})}}^{\sharp -1}(T^{}_{_{X({u})}} C_{u})
\cap N^*_{_{X({u})}}(C_{u}),\quad {\rm for}\ {u}=0,\pi.$$

Note that a reparametrization of the path
$\sigma\mapsto\sigma'=\sigma+\delta\sigma$ with $\delta\sigma(u)=0,
u=0,\pi$ corresponds to a gauge symmetry with
$\epsilon_k=\psi_{k\sigma}\delta\sigma$. One may also check that as
for the free BC or the coisotropic case the characteristic
distribution of $\Omega$ has finite codimension.

As discussed in Section 2 the presymplectic structure induces a Poisson
algebra ${\cal P}$ in the phase space $P(M,C_0,C_\pi)$. On the other hand,
we have Poisson algebras in $C_0$ and $C_\pi$. We turn now to study the
relation between them. 

We first analyse under which circumstances a function
$F(X,\psi)=f(X(0))$, $f\in{C}^\infty(M)$ belongs to ${\cal P}$, i.e. when
it has a hamiltonian vector field $\delta_F$.  Solving the
corresponding equation we see that the general solution is of the form
(\ref{Hamsymm}) with
\begin{equation}\label{cond0}
\epsilon(0)-df_{_{X(0)}}\in
N^*_{_{X(0)}}(C_0),
\qquad\epsilon(0)\in \Gamma_{_{X(0)}}^{\sharp -1}(T^{}_{_{X(0)}} C_0)
\end{equation}
and  
$$\epsilon(\pi)\in
\Gamma_{_{X(\pi)}}^{\sharp -1}(T^{}_{_{X(\pi)}} C_\pi)
\cap N^*_{_{X(\pi)}}(C_\pi).$$
We saw in Section 3 (see Theorem 3) that assuming ${\rm
dim}(\Gamma^\sharp(N^*_p(C_0))+T_pC_0)=const.$, equation (\ref{cond0})
can be solved in $\epsilon(0)$ if and only if $F$ is a gauge invariant
function (i.e. it is invariant under (\ref{Hamsymm})).  This is
equivalent to saying that $f+{\cal I}_0$ belongs to the Poisson algebra
${\cal C}(\Gamma,M,C_0)$.  (Here ${\cal I}_0$ is the ideal of functions that
vanish on $C_0$).

Now, given two such functions
$F_1$ and $F_2$ associated to 
$f_1+{\cal I}_0,f_2+{\cal I}_0 \in{\cal C}(\Gamma,M,C_0)$
and with gauge field $\epsilon_1$ and $\epsilon_2$ respectively,
one immediately computes the Poisson bracket 
$\{F_1,F_2\}_P=\Omega(\delta_{F_1},\delta_{F_2})$ to give
\begin{equation}
\{F_1,F_2\}_P=\Gamma^{ij}\epsilon_{1i}(0)\epsilon_{2j}(0)
\end{equation}
This coincides with the restriction to $C_0$ of
$\{f_1+{\cal I}_0,f_2+{\cal I}_0\}_{C_0}$ and defines a Poisson homomorphism
between ${\cal C}(\Gamma,M,C_0)$ and the Poisson algebra of
$P(M,C_0,C_\pi)$. This homomorphism is $J_0^*$, the pull-back defined
by $J_0$, and the latter turns out to be a Poisson map. In an
analogous way we may show that $J_\pi$ is an anti-Poisson map and
besides
$$\{f_0\circ J_0, f_\pi\circ J_\pi\}=0\quad {\rm for\ any}\  
f_{u}\in{\cal C}(\Gamma,M,C_{u}),\ {u}=0,\pi.$$

The previous considerations 
can be summarized in the following diagram
\begin{equation}
\begin{matrix}
&\displaystyle
J^*_0
&\displaystyle
&\displaystyle
J^*_\pi
&\displaystyle
\cr
{\cal C}(\Gamma,M,C_0)
&\displaystyle
\longrightarrow
&\displaystyle
{\cal P}
&\displaystyle
\longleftarrow
&\displaystyle
{\cal C}(\Gamma,M,C_\pi)
\cr
\end{matrix}
\end{equation}
in which $J_0^*$ is a Poisson homomorphism,
$J_\pi^*$ antihomomorphism and
the image of each map is the commutant 
(with respect to the Poisson bracket)
of the other. In particular it implies that the reduced phase space is finite-dimensional, as claimed above.

This can be considered as a generalization 
of the symplectic dual pair to the context of Poisson algebras. 
 
\section{Conclusions}

We have generalized the results of \cite{CaFe3} to allow for 
non coisotropic branes in the Poisson-Sigma model.
 
In this more general situation we have to consider the 
reduction of Poisson brackets to a submanifold $C$ of the original
Poisson manifold $M$. This is achived in a canonical way at
the price of ending up with a Poisson algebra on 
a subset of ${C}^\infty(C)$ rather than
a Poisson structure on $C$. In more physical terms
we can rephrase the previous considerations
by saying that we are led to select certain 
observables on $C$ which are the functions
on the constrained phase space that belong to the Poisson algebra.

Two cases of interest are either when these observables fill
${C}^\infty(C)$ or when they are the functions invariant under the
gauge transformations generated by the (first-class) constraints.  We
show that the constant rank of the Poisson bracket of (a local regular
basis of) constraints is a sufficent conditions for having one case or
the other. The Poisson bracket in these situations is the Dirac
bracket (for gauge invariant functions in the second case).  We show
that in this setup it is possible to determine consistent BC for the
Poisson-Sigma model with the base map at the boundary taking values in
$C$.  The resulting Poisson-Sigma model enjoys the basic properties of
a topological theory (at the classical level), namely the
characteristic distribution has finite codimension (the phase space
reduced by the symmetries is finite-dimensional), and the
reparametrization of the paths is among the gauge symmetries.

Quantization of the theory requires the introduction
of ghost fields and BC for them. We have checked
that in the constant rank case, BC for ghost fields 
can be chosen so that they are consistent with BRST symmetry
at the boundary.

Feynman expansion of certain Green's functions of the Poisson-Sigma
model on the disc, with free BC, gives rise to the
formal deformation quantization of Poisson brackets found by
Kontsevich \cite{Kon}. It is natural to ask for the corresponding
calculation with non-free BC. The coisotropic case
has been worked out in \cite{CaFe3} leading, under some suplementary
assumptions, to a 
deformed associative product in the space of quantum-gauge
invariant functions on $C$.
In our case we would
expect some similar result but with the Dirac bracket playing the
relevant role.
%this is at least what one gets when performing
%the formal integration of field $\psi$ in the symplectic case.
However, this is a subtle issue: first note that in the Dirac bracket
the inverse of the matrix of the Poisson brackets for second-class
constraints appears. This inverse cannot be obtained by the standard
perturbation theory of refs. \cite{CaFe},\cite{CaFe3} around the zero
Poisson bracket. The obstruction can be also seen from the fact that
the propagator (in the standard perturbation theory) for the modes
corresponding to the second-class constraints does not exist.  A way
out of this situation could be to redefine the perturbation theory by
integrating out the fields $\psi_A$ (with the notation of section 5,
(\ref{localcoor})) using the fact that $\Gamma^{AB}$ is in this case
invertible.  The details of this computation will be the subject of
further research.

In the case of a manifold $[0,\pi]\times{\mathbb R}$ with two connected
components we analyse the relation between the BC at
both components. We show that only points in the interesection of the
two branes with a common symplectic leaf of $M$ appear as boundary
points in the classical solutions. If the evaluation of solutions at
the boundary points $J_0$ and $J_\pi$ are surjective submersion onto
the corresponding branes $C_0$ and $C_\pi$, we may show that $J_0$
is a Poisson map while $J_\pi$ is anti-Poisson. Furthermore, the
pull-back by $J^*_0$ of the Poisson algebra associated to $C_0$ is the
commutant of the pull-back by $J^*_\pi$ and viceversa. This defines a
dual pair structure in the context of Poisson algebras that
generalizes the concept of symplectic dual pair of \cite{Weinstein}.

When the target is a Poisson-Lie group (\cite{Falceto},\cite{CalFal}),
the question of dual BC raises naturally. In ref. \cite{CalFal}
it is shown that free boundary conditions are related by
duality. A more general study was attempted in (\cite{BonZab}) 
in the coisotropic case. It would be
interesting to address the problem in the context of more general BC.

The quantum counterpart of the strip was studied in 
\cite{CaFe3} in the coisotropic case. There, the authors found 
a bimodule structure with the quantum algebras associated to
$C_0$ ($C_\pi$) acting by the deformed product on the right 
(left) on the algebra associated to $C_0\cap C_\pi$. It is then 
natural to ask for the generalization of these results
for the case of more general BC.

\vskip 4mm
\noindent{\bf Acknowledgments:}
 We benefited from discussions with M. Asorey and J. F. Cari\~nena.
Special thanks are due to K. Gaw\c{e}dzki for private correspondence
sharing valuable insights at the early stage of this paper. 
We also acknowledge the referee for a careful reading and useful suggestions.
This work was supported by MCYT (Spain), grant FPA2003-02948. I. C. 
is supported by MEC (Spain), grant FPU.

\end{document}